\documentclass[aps, pre, showkeys, showpacs, reprint, twocolumn, nofootinbib, superscriptaddress, floatfix]{revtex4-1}

\usepackage[utf8]{inputenc}
\usepackage[T1]{fontenc}
\usepackage{amsmath}
\usepackage{graphicx}
\usepackage{psfrag}
\usepackage{lipsum}
\usepackage{color}

\begin{document}

\title{Game of collusions}

\author{Krzysztof Malarz}
\homepage{http://home.agh.edu.pl/malarz/}
\email{malarz@agh.edu.pl}
\affiliation{\href{http://www.agh.edu.pl/}{AGH University of Science and Technology},
\href{http://www.pacs.agh.edu.pl/}{Faculty of Physics and Applied Computer Science},\\
al. Mickiewicza 30, 30-059 Krakow, Poland.}

\author{Krzysztof Ku{\l}akowski}
\email{kulakowski@fis.agh.edu.pl}
\affiliation{\href{http://www.agh.edu.pl/}{AGH University of Science and Technology},
\href{http://www.pacs.agh.edu.pl/}{Faculty of Physics and Applied Computer Science},\\
al. Mickiewicza 30, 30-059 Krakow, Poland.}

\begin{abstract}
A new model of collusions in an organization is proposed.
Each actor $a_{i=1,\cdots,N}$ disposes one unique good $g_{j=1,\cdots,N}$. 
Each actor $a_i$ has also a list of other goods which he/she needs, in order from desired most to those desired less.
Finally, each actor $a_i$ has also a list of other agents, initially ordered at random.
The order in the last list means the order of the access of the actors to the good $g_j$.
A pair after a pair of agents tries to make a transaction.
This transaction is possible if each of two actors can be shifted upwards in the list of actors possessed by the partner.
Our numerical results indicate, that the average time of evolution scales with the number $N$ of actors approximately as $N^{2.9}$.
For each actor, we calculate the Kendall's rank correlation between the order of desired goods and actor's place at the lists of the good's possessors.
We also calculate individual utility funcions $\eta_i$, where goods are weighted according to how strongly they are desired by an actor $a_i$, and how easily they can be accessed by $a_i$.
Although the individual utility functions can increase or decrease in the time course, its value averaged over actors and independent simulations does increase in time.
This means that the system of collusions is profitable for the members of the organization. 
\end{abstract}



\maketitle

\section{Introduction}

An ambitious and careful program of social research on organizations, known as Strategic Analysis (SA), has been framed nearly 40 years ago by Crozier and Friedberg \cite{Crozier}. 
The key idea of SA is that the structure and activity in organizations are shaped by the relations of power. Further, an actor is autonomous, hence her/his behaviour is uncertain and this uncertainty is an ingredient part of the actor's power over other actors. Further, each organization is contingent; it comes out as an artificial construct to solve a given problem. Therefore any theory which tries to derive a characteristics of an organizations from 'natural' conditions is doomed to failure. These frames make SA applicable to qualitative discussions of specific case studies as of Airbus \cite{ap1}, Lidl \cite{ap2}, or selected (unnamed) universities \cite{3u}, but particularly difficult to be translated into terms of a formal model of some generality. Also, any such translation contains elements which are necessarily arbitrary.

Yet, only recently the task has been undertaken in a few directions. Particular configurations of the relations of power in an organization (principal, supervisor, agent) have been investigated by Vafa\"i \cite{va1,va2}, who analysed couplings between different kinds of collusions in terms of mathematical theorems. The approach by Sibertin-Blanc and coworkers \cite{s1,s2,s3} develops the frames of SA in a consistent and methodical way. In this approach, each agent (actor) controls some relations which other agents can profit by. Also, each agent distributes his own stakes over the relations he/she can profit by. As an outcome of this game, played by all agents simultaneously, they get their capacities, which depend on both the relations and the stakes. In Ref.~\cite{s3}, the algorithm---termed as SocLab---is applied to the case of management of floods of the Touch river in France.

The aim here is to analyse in detail one particular aspect of SA, i.e. binary collusions. By a collusion we mean, after Ref.~\cite{mw}, {\it secret agreement or cooperation especially for an illegal or deceitful purpose}. We are not interested, however, in specifying to what extent the collusions to be simulated are illegal or even deceitful. We are going to concentrate on the process of establishing a hierarchy in an organization. Similarly to the papers reported above, our motivation is to refer to the frames of SA, as described by Crozier and Friedberg \cite{Crozier}. Yet, it seems to us that the process of setting of hierarchies (ST) is at least as close to SA as the algorithm SocLab \cite{s3}. In particular:
\begin{itemize}
\item as stated in Ref.~\cite{Crozier}, there are always some relations of power in an organization; a simulation of ST should reproduce their dynamics;
\item as stated in Ref.~\cite{Crozier}, an organization itself is a social construct; so is the hierarchy, constructed individually by each actor;
\item the actors are of limited rationality and limited information; therefore they do not plan their behavior in long time scale, just---as also stated in Ref.~\cite{Crozier}---catch occassions;
\item for the same reason, actors can neither predict nor control the behaviors of other actors, as those maintain their spheres of uncertainty. A successful cooperation with one actor does not preclude its future breaking for the sake of a more profitable cooperation with another actor;
\item a game composed of binary subgames seems more realistic than a collective reorientations of relations, with perfect information available immediately for all agents.
\end{itemize}

Last but not least, as it will be shown below, the only parameter of our model is the number of agents. Obviously, our approach is rather explanatory than predictive. Yet it seems to us worthwhile to try to understand a selected process of SA, even if its description remains separated from particular case studies.

Our scenario is as follows. In a set of agents, each agent disposes some kind of resources. Also, each agent needs resources of all other kinds, with his/her individual order of needs. Finally, each agent has a list of all other agents, with their hierarchy equivalent to their order in the list. Once two agents simultaneously find that it would be profitable to be advanced in the hierarchy of the other agent, they shift each other upwards by one position. Then, the lists of hierarchies are the only elements which evolve.

In the next section~\ref{sec:model}, our algorithm is explained in details.
The Sec.~\ref{sec:results} presents results of computer simulations.
Finally, Sec.~\ref{sec:disc} contains discussion of the results and conclusions.

\section{\label{sec:model}Model}

Let $\mathbf{A}=\{a_1,\cdots,a_N\}$ and $\mathbf{G}=\{g_1,\cdots,g_N\}$ denote sets of $N$ agents and $N$ goods, respectively.
Every agent $i$ posses single and unique good $\sigma_i\in\mathbf{G}$, i.e.
\begin{equation}
(\sigma_1,\sigma_2,\cdots,\sigma_N)=\mathcal{P}(\mathbf{G}),
\label{eq:1}
\end{equation}
where $\mathcal{P}(\mathbf{G})$ denotes a random permutation of set $\mathbf{G}$.
Every agent $i$ desires all other $(N-1)$ goods in a given order
\begin{equation}
\vec\xi_i=\mathcal{P}(\mathbf{G}_i),
\label{eq:2}
\end{equation}
where $\mathcal{P}(\mathbf{G}_i)$ is a random permutation of set $\mathbf{G}_i=\mathbf{G}\setminus\{\sigma_i\}$.
A sequences $\vec\xi_i$ do not change during simulation.
Every agent $i$ orders sequentially all other $(N-1)$ agents.
This order is represented as a sequence $\vec\zeta_i^{\,t}$ which may evolve during time $t$.
Initially, the list $\vec\zeta_i^{\,0}$ is chosen randomly, i.e.
\begin{equation}
\vec\zeta_i^{\,0}=\mathcal{P}(\mathbf{A}_i),
\label{eq:3}
\end{equation}
where $\mathcal{P}(\mathbf{A}_i)$ stands for a random permutation of set $\mathbf{A}_i=\mathbf{A}\setminus\{a_i\}$.

In each time step $t$ a pair $(a_m,a_n)$ of agents is selected randomly.
Let $k_m$ denotes position of agent $a_m$ in the list $\vec\zeta_n^{\,t-1}$ and $k_n$ denotes position of agent $a_n$ in the list $\vec\zeta_m^{\,t-1}$.
If simultaneously, item $(k_m-1)$ on the list $\vec\zeta_n^{\,t-1}$ and item $(k_n-1)$ on the list $\vec\zeta_m^{\,t-1}$ represent agents which have goods considered by agents $a_m$ and $a_n$ as less valuable than goods of agents in positions $k_m$ and $k_n$, then {\em transaction} takes place.
During transaction agents in positions $(k_m-1)$ and $k_m$ on the list $\vec\zeta_n^{\,t}$ are swapped.
The same procedure is realised on the $\vec\zeta_m^{\,t}$ list, i.e. agents in positions $(k_n-1)$ and $k_n$ are swapped as well.
We call promoted and demoted agents `winners' and `losers' of the transactions, respectively.

The results are averaged over $M$ independent simulations, i.e.
various sequences of $(\sigma_1,\sigma_2,\cdots,\sigma_N)$ [Eq.~\eqref{eq:1}],
various hierarchies of agents goods $\vec\xi_i$ $(i=1,\cdots,N)$ [Eq.~\eqref{eq:2}]
and various initial sequences $\vec\zeta_i^{\,0}$ [Eq.~\eqref{eq:3}].

\begin{figure}
\psfrag{N=6}[c]{(a) $N=6$}
\psfrag{N=12}[c]{(b) $N=12$}
\psfrag{N=24}[c]{(c) $N=24$}
\psfrag{t}{$t$}
\psfrag{i}{$i$}
\psfrag{ken}{$\kappa_i$}
\psfrag{kendall}[c]{$\kappa_i$}
\includegraphics[width=.99\columnwidth]{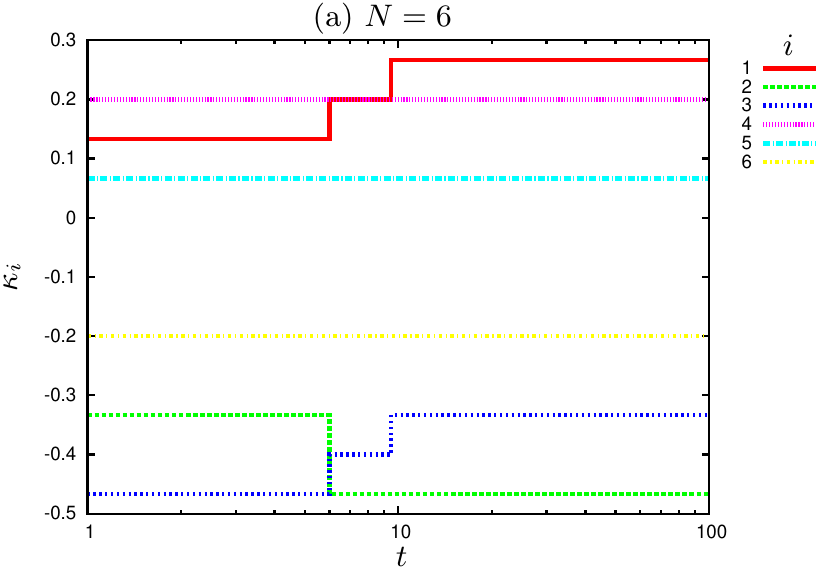}\\[1mm]
\includegraphics[width=.99\columnwidth]{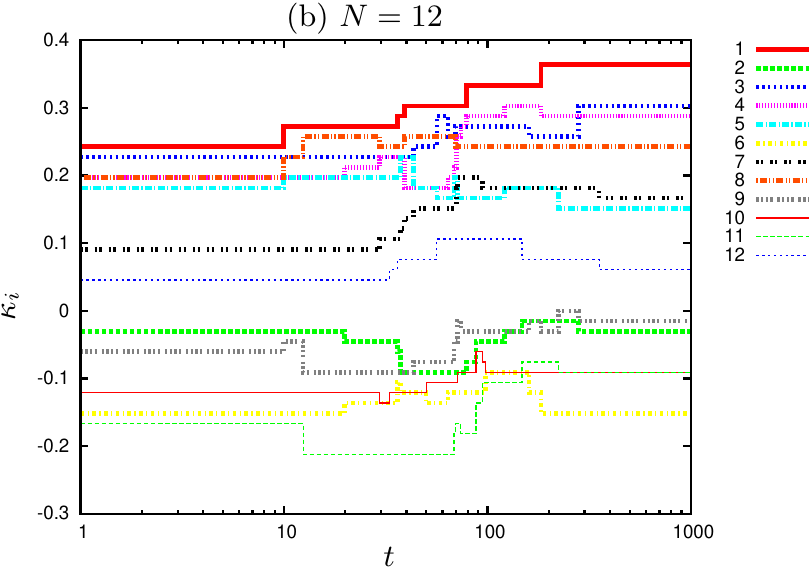}\\[1mm]
\includegraphics[width=.99\columnwidth]{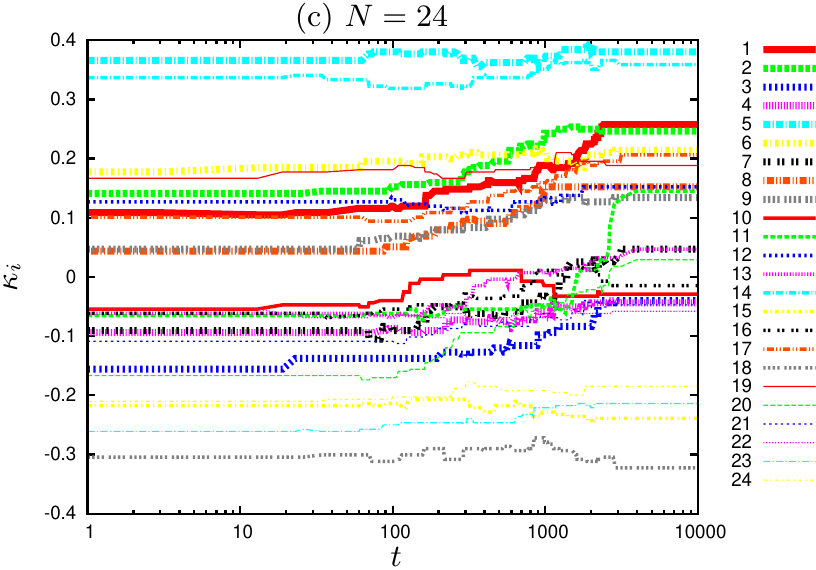}
\caption{\label{fig:kappa}Time evolution of the Kendall's rank correlation coefficient $\kappa$ for various system sizes $N=6$, 12, 24 from top to bottom.}
\end{figure}

\begin{figure*}
\psfrag{rho}{$\rho$}
\psfrag{w}{$w$}
\psfrag{l}{$l$}
\psfrag{D k}[c]{$\Delta\kappa_i^{w,l}$}
\psfrag{N=6, Nrun=1e5}[c]{(a) $N=6$, $M=10^5$}
\psfrag{N=12, Nrun=1e5}[c]{(b) $N=12$, $M=10^5$}
\psfrag{N=24, Nrun=1e5}[c]{(c) $N=24$, $M=10^5$}
\psfrag{N=48, Nrun=1e3}[c]{(d) $N=48$, $M=10^3$}
\psfrag{N=96, Nrun=1e2}[c]{(e) $N=24$, $M=10^3$}
\includegraphics[width=0.85\textwidth]{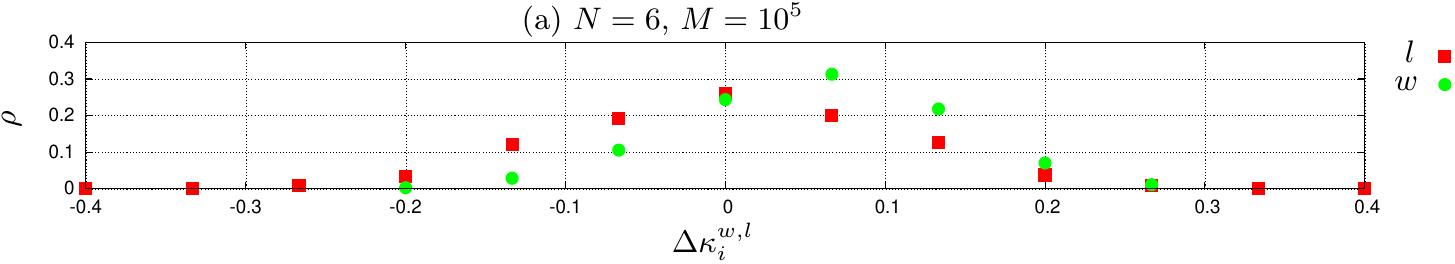}\\[1mm]
\includegraphics[width=0.85\textwidth]{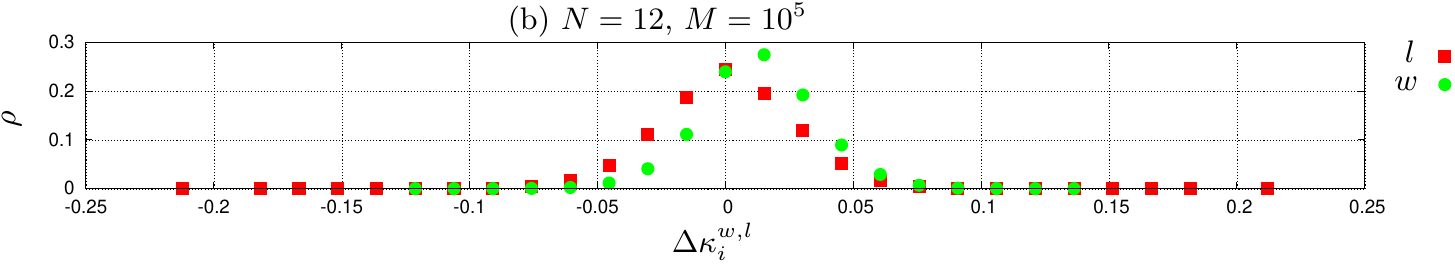}\\[1mm]
\includegraphics[width=0.85\textwidth]{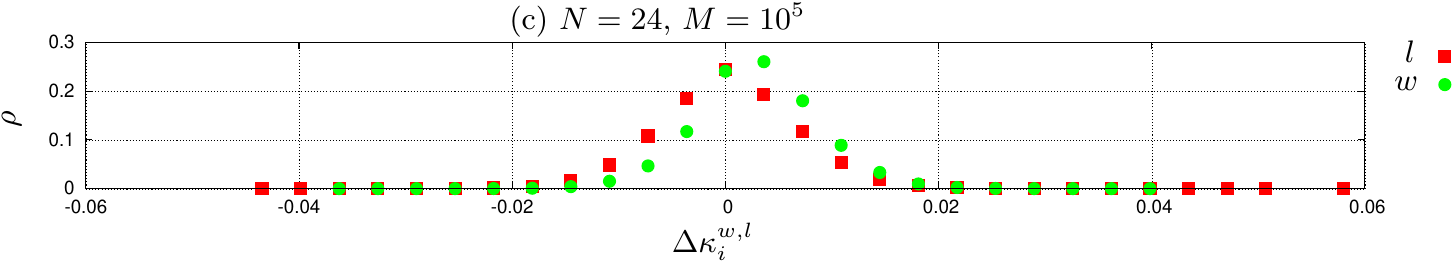}\\[1mm]
\includegraphics[width=0.85\textwidth]{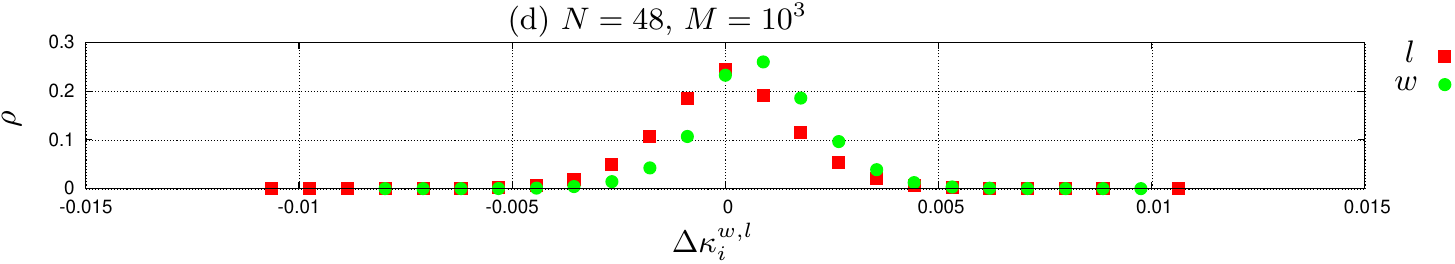}\\[1mm]
\caption{\label{fig:hisdeltakendall}Probability distributions $\rho(\Delta\kappa_i^{w,l})$ of Kendall's statistics changes during transactions $\Delta\kappa_i^{w,l}=\kappa_i(t+1)-\kappa_i(t)$.
The circles and squares correspond to transactions `winners' ($w$) and `losers' ($l$), respectively.
}
\end{figure*}

Basically, the simulation is performed by subsequent transactions between randomly selected pairs of actors, if the transaction is profitable for both. We note however that the system evolution is equivalent to a probabilistic cellular automaton \cite{wolf}, where the rule is applied simultaneously to all the pairs, but the mutually profitable transactions are performed with some small probability $\pi$. The condition of this equivalence is that $\pi$ is taken so small as to exclude two simultaneous transactions of the same actor in the same time step.

\subsection{\label{sec:example}An example of model rules application}
To illustrate the model's rules let us consider a group of $N=6$ agents with their goods $\sigma_i$, hierarchy of goods $\vec\xi_i$ and initial hierarchy of agents $\vec\zeta_i^{\,0}$:

\begin{equation}
\begin{array}{rc}
a_ 1: \sigma_1 = g_1, \\
\vec\xi_1= 	& (g_6, g_5, g_3, g_4, g_2),\\
\vec\zeta_1^{\,0}= 	& (a_2, a_6, a_5, a_4, a_3),\\
a_2:  \sigma_2 = g_6,\\
\vec\xi_2= 	& (g_2, g_1, g_5, g_4, g_3),\\
\vec\zeta_2^{\,0}= 	& (a_5, a_4, a_1, a_6, a_3),\\
a_3: \sigma_3 = g_5,\\
\vec\xi_3= 	& (g_1, g_4, g_3, g_2, g_6),\\
\vec\zeta_3^{\,0}= 	& (a_2, a_4, a_5, a_1, a_6),\\
a_4: \sigma_4 = g_2,\\
\vec\xi_4= 	& (g_5, g_4, g_6, g_1, g_3),\\
\vec\zeta_4^{\,0}=	& (a_2, a_5, a_1, a_6, a_3),\\
a_5: \sigma_5 = g_3,\\
\vec\xi_5= 	& (g_5, g_2, g_6, g_1, g_4),\\
\vec\zeta_5^{\,0}= 	& (a_6, a_4, a_3, a_1, a_2),\\
a_6: \sigma_6 = g_4,\\
\vec\xi_6= 	& (g_2, g_6, g_5, g_1, g_3),\\
\vec\zeta_6^{\,0}= 	& (a_2, a_1, a_5, a_3, a_4).\\
\end{array}
\label{eq:t=0}
\end{equation}

\begin{itemize}
\item Let us assume that in $t=1$ the pair $(a_2,a_3)$ is selected.
In such situation a transaction will not take place as agent $a_2$ is already at the first position on $\vec\zeta_3^{\,0}$ list and thus for every agent $\vec\zeta_i^{\,1}=\vec\zeta_i^{\,0}$ ($i=1,\cdots,6$).
Please note however, that simultaneously the agent $a_2$ would like to promote agent $a_3$ and would like to demote agent $a_6$, as $\sigma_3=g_5$ is higher (third) than $\sigma_6=g_4$ (fourth) in agent $a_2$'s hierarchy of goods $\vec\xi_2$.

\item Now, in $t=2$, let a pair $(a_4,a_6)$ is considered.
Agent $a_6$ is located in $\vec\zeta_4^{\,1}$ list directly after $a_1$ while agent $a_3$ is directly before $a_4$ in $\vec\zeta_6^{\,1}$ list.
Agent $a_4$ would like to promote agent $a_6$ and to demote agent $a_1$, as $\sigma_6=g_4$ is higher (second) than $\sigma_1=g_1$ (fourth) in agents $a_4$'s hierarchy of goods $\vec\xi_4$.
Agent $a_6$ would like to promote agent $a_4$ and to demote agent $a_3$, as $\sigma_4=g_2$ is higher  (first) than $\sigma_3=g_5$  (third) in agents $a_6$'s hierarchy of goods $\vec\xi_6$.
As both agents $a_4$ and $a_6$ would like to promote each other, the transaction takes place, what modifies the lists $\vec\zeta_4^{\,2}$ and $\vec\zeta_6^{\,2}$:
\begin{equation}
\begin{array}{rc}
a_4:\\ 
\vec\zeta_4^{\,2}= & (a_2, a_5, \textcolor{green}{a_6}, \textcolor{red}{a_1}, a_3),\\
a_6:\\ 
\vec\zeta_6^{\,2}= & (a_2, a_1, a_5, \textcolor{green}{a_4}, \textcolor{red}{a_3}).\\
\label{eq:t=2}
\end{array}
\end{equation}
All other lists remain unchanged, i.e. $\vec\zeta_i^{\,2}=\vec\zeta_i^{\,1}$ for $1\le i\le 6$ and $i\notin \{4,6\}$.
Agents $a_4$ and $a_6$ are the `winners' of the transaction, while `losers' $a_1$ and $a_3$ are demoted during this transaction.

\item In the next time step a pair $(a_5,a_1)$ is selected.
Agent $a_1$ would like to promote agent $a_5$ and to demote agent $a_6$, as $\sigma_5=g_3$ is higher (third) than $\sigma_6=g_4$ (fourth) in agents $a_1$'s hierarchy of goods $\vec\xi_1$.
However, agent $a_5$ does not want to promote agent $a_1$, as $\sigma_1=g_1$ is lower (fourth) than $\sigma_3=g_5$ (first) in agents $a_5$'s hierarchy of goods $\vec\xi_5$.
In this time step the transaction is absent.

\item In the fourth time step a pair $(a_5,a_3)$ is selected.
An agent $a_4$ is present in agent $a_5$ list $\vec\xi_5$ directly before $a_3$ and simultaneously he/she is directly before agent $a_5$ on $\vec\xi_3$ list.
The good of agent $a_4$ ($\sigma_4=g_2$) is less valuable for $a_3$ than good $\sigma_5=g_3$ and it is less noteworthy than good $\sigma_3=g_5$ for agent $a_5$.
Thus agent $a_4$ will be demoted on both lists and he/she will become a `double-looser' of this transaction.
\begin{equation}
\begin{array}{rc}
a_3:\\ 
\vec\zeta_3^{\,4}= & (a_2, \textcolor{green}{a_5}, \textcolor{red}{a_4}, a_1, a_6),\\
a_5:\\ 
\vec\zeta_5^{\,4}= & (a_6, \textcolor{green}{a_3}, \textcolor{red}{a_4}, a_1, a_2).\\
\label{eq:t=4}
\end{array}
\end{equation}
All other list remain unchanged, i.e. $\vec\zeta_i^{\,4}=\vec\zeta_i^{\,3}$ for $1\le i\le 6$ and $i\notin\{3,5\}$, etc...
\end{itemize}

\begin{figure}
\psfrag{eta}{$\eta_i$}
\psfrag{t}{$t$}
\psfrag{i}{$i$}
\psfrag{N=6}[c]{(a) $N=6$, $\eta_i^{\min}\approx 0.351$, $\eta_i^{\max}\approx 1.405$}
\psfrag{N=12}[c]{(b) $N=12$, $\eta_i^{\min}\approx 0.0521$, $\eta_i^{\max}\approx 1.669$}
\psfrag{N=24}[c]{(c) $N=24$, $\eta_i^{\min}\approx 0.000833$, $\eta_i^{\max}\approx 1.706$}
\includegraphics[width=.99\columnwidth]{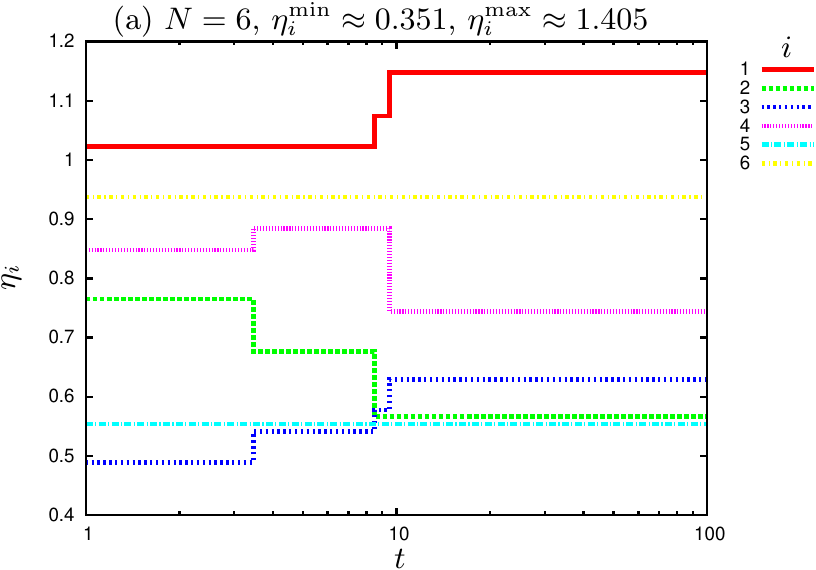}\\[1mm]
\includegraphics[width=.99\columnwidth]{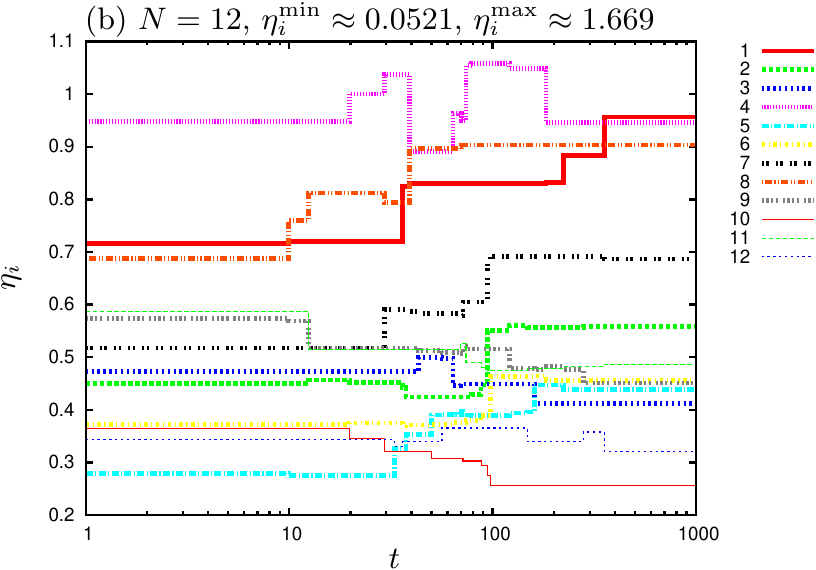}\\[1mm]
\includegraphics[width=.99\columnwidth]{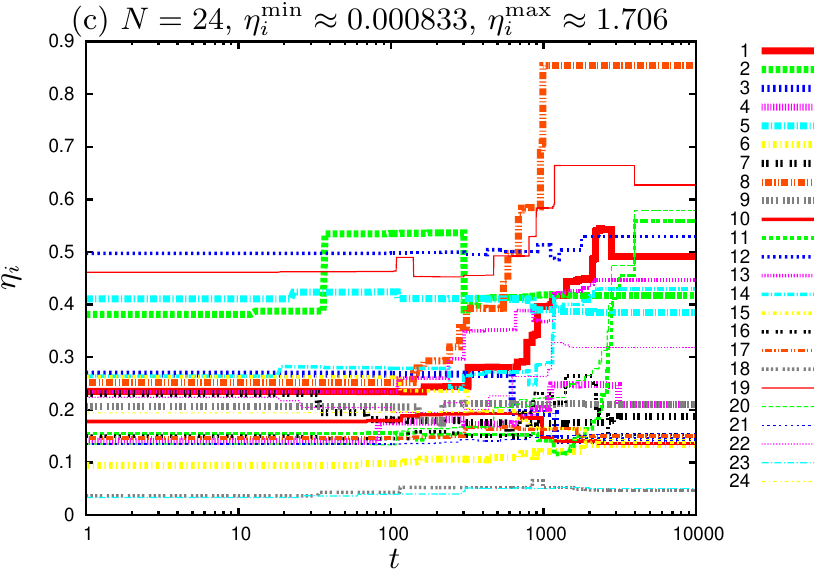}
\caption{\label{fig:etavst}Time evolution of the parameters $\eta_i(t)$ for $N=6$, 12 and 24 agents.
The ranges of possible $\eta_i\in[\eta_i^{\min},\eta_i^{\max}]$ values are indicated in sub-figures headlines.}
\end{figure}

\subsection{\label{sec:defkappa}The Kendall's rank correlation coefficient.}

Technically, in order to calculate Kandall's $\tau$ coefficient (in this paper denoted as $\kappa$) one have to check all possible pairs of $(x_i,x_j)$ and $(y_i,y_j)$ of sequences $\vec x=(x_1, x_2, \cdots, x_N)$ and $\vec y=(y_1, y_2, \cdots, y_N)$. 
Pairs $(x_i,x_j)$ and $(y_i,y_j)$ are called concordant when $x_i<x_j$ and $y_i<y_j$ (or $x_i>x_j$ and $y_i>y_j$), while they are called discordant if $x_i<x_j$ and $y_i>y_j$ (or $x_i>x_j$ and $y_i<y_j$).

The Kendall's $\kappa$ coefficient \cite{Kendall} is defined as
\begin{equation}
\kappa\equiv\frac{2(p-q)}{N(N-1)},
\label{eq:Kendall}
\end{equation}
where $p$ and $q$ are numbers of concordant and discordant pairs among ${N \choose 2}$ available pairs of $N$-items long sequences $\vec x$ (and/or $\vec y$).

The value of $\kappa=+1$ means that all pairs $(x_i,x_j)$ and $(y_i,y_j)$ are concordant and $\kappa=-1$ only when all pairs $(x_i,x_j)$ and $(y_i,y_j)$ are discordant.

\subsection{\label{sec:defeta}The utility coefficient $\eta$.}

\begin{figure}
\psfrag{N=12, i=1}[c]{(a) $N=12$, $i=1$}
\psfrag{N=12, i=2}[c]{(b) $N=12$, $i=2$}
\psfrag{t}{$t$}
\psfrag{eta}{$\eta_i$}
\psfrag{ken}{$\kappa_i$}
\psfrag{kendall}[c]{$\kappa_i$}
\includegraphics[width=.90\columnwidth]{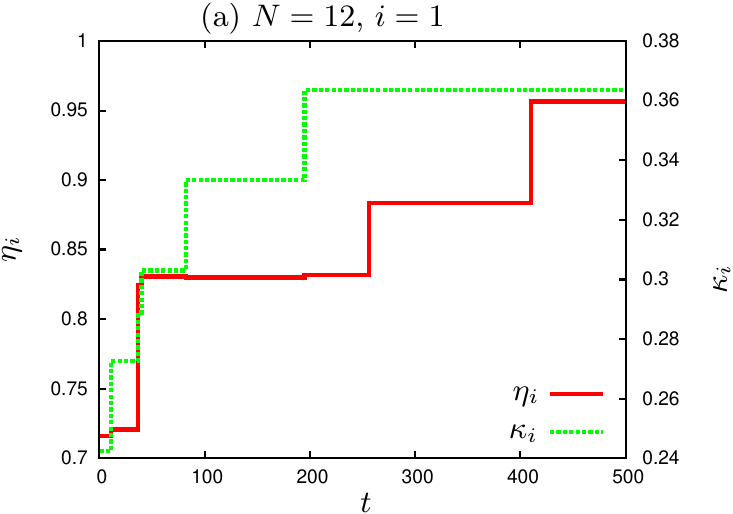}\\[1mm]
\includegraphics[width=.90\columnwidth]{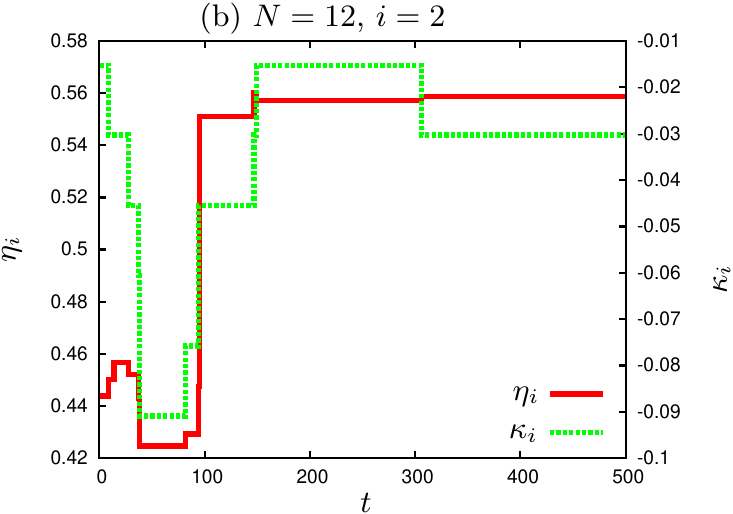}
\caption{\label{fig:kappavseta}The comparison of time evolution of parameters $\kappa_i(t)$ (values on right axis) and $\eta_i(t)$ (values on left axis) for two agents (a) $a_1$ and (b) $a_2$ among group of a dozen of agents.}
\end{figure}

We will show later (Sec.~\ref{sec:eta}, Fig.~\ref{fig:kappavseta}) that Kendall's coefficient changes are not sensitive for all kinds of transaction results.
Thus we propose another utility coefficient
\begin{equation}
\eta_i(t)\equiv\sum_{j=1}^{N-1}\alpha^{-\frac{1}{2}\left[j+\nu_{ij}(t)\right]},
\label{eq:etai}
\end{equation}
where $\alpha\equiv 2$, $j$ marks position of the $j$-th good on the list $\xi_i$, while $\nu_{ij}(t)$ denotes $i$-th agent position in time $t$ on the list $\vec\zeta_k^{\,t}$, where $k$ is the label of the agent $a_k$ having good being $j$-th on $\vec\xi_i$ list.

For example, if the initial conditions are given by Eq.~\eqref{eq:t=0} then $\nu_{1j}(t=0)$ are 3, 4, 4, 2, 3 and will become 3, 4, 4, 2, 4 [$\nu_{1j}(t=2)$] after transaction described in Eq.~\eqref{eq:t=2} for $j=1,2,\cdots,5$, respectively.

In contrast to the Kendall's rank correlation coefficient $-1\le\kappa\le+1$ only positive values of $\eta_i$ are possible.
The maximal available value of $\eta_i$
\begin{equation}
\eta_i^{\max}=\sum_{j=1}^{N-1} 2^{-(j+1)/2} = \left(1+\frac{1}{\sqrt{2}}\right) (1-2^{1/2-N/2})
\label{eq:etamax}
\end{equation}
corresponds to the situation when agent $a_i$ is in the first position on all $\zeta_{j\ne i}$ lists, i.e. $\nu_{ij}=1$ for $j=1,2,\cdots,N-1$.
And occurrence of the minimal available value of $\eta_i$
\begin{equation}
\eta_i^{\min}=\sum_{j=1}^{N-1} 2^{-(j+N-1)/2} = \frac{2^{1-N}\cdot(\sqrt{2}-2^{N/2})}{\sqrt{2}-2}
\label{eq:etamin}
\end{equation}
means that the agent $a_i$ is in the last $(N-1)$-th position on all $\zeta_{j\ne i}$ lists, i.e. $\nu_{ij}=N-1$ for $j=1,2,\cdots,N-1$.
Thus the parameter $\eta_i(t)$ may measure the `prestige' of agent $a_i$.

\section{\label{sec:results}Results.}

\begin{figure*}
\psfrag{<eta>}{$\eta$}
\psfrag{N=12}{(a) $N=12$}
\psfrag{N=24}{(b) $N=24$}
\psfrag{N=48}{(c) $N=48$}
\psfrag{N=96}{(d) $N=96$}
\psfrag{t}{$t$}
\psfrag{ave}{$\langle\eta\rangle$}
\includegraphics[width=.95\columnwidth]{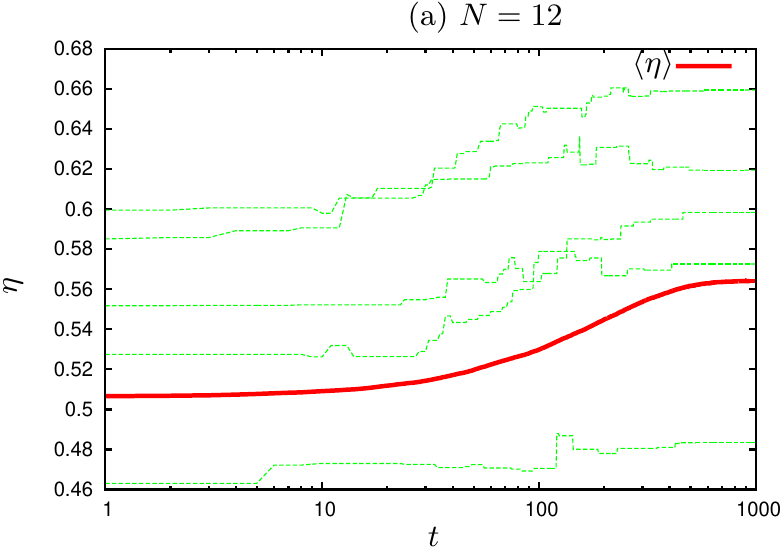}
\includegraphics[width=.95\columnwidth]{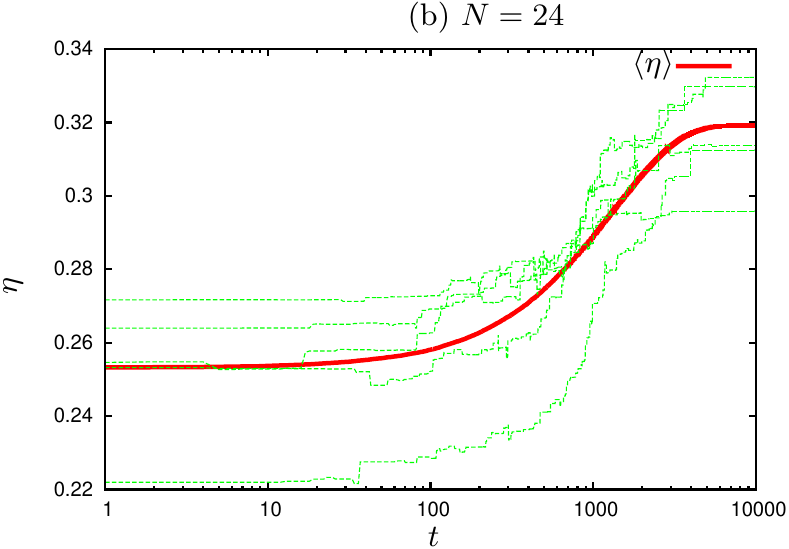}\\
\includegraphics[width=.95\columnwidth]{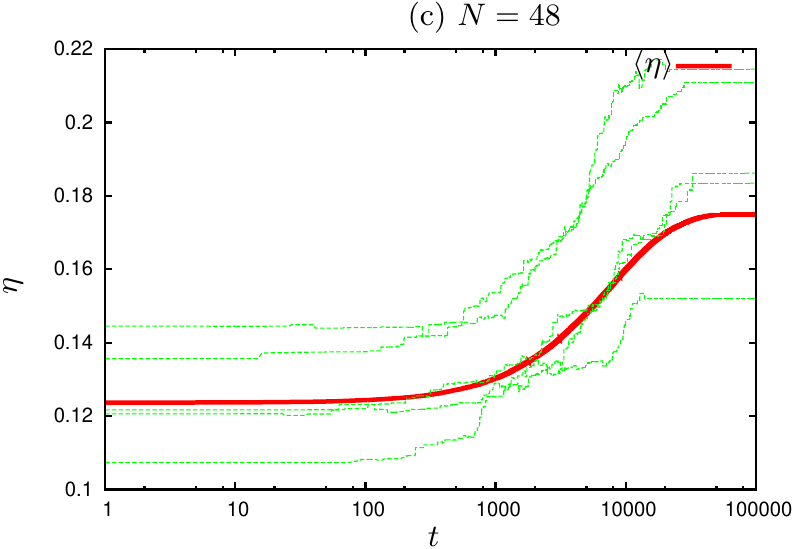}
\includegraphics[width=.95\columnwidth]{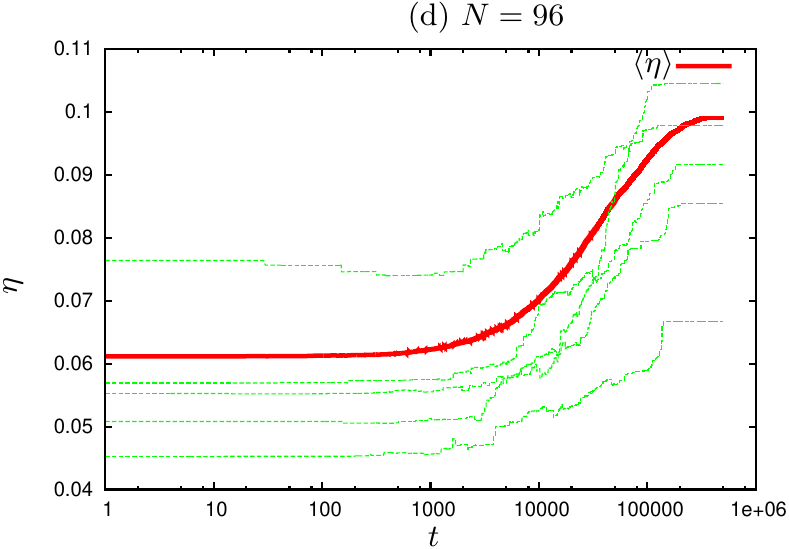}
\caption{\label{fig:aveta}The time evolution of an average parameter $\eta\equiv\bar\eta_i$, where bar over symbol stands for an average over all $N$ agents, and $\langle\cdots\rangle$ is an additional averaging over $M$ independents simulations.
The solid red thick line corresponds to time evolution of $\langle\eta(t)\rangle$, while thin lines presents examples of $\eta(t)$ evolutions in a single running.
The values of $\langle\eta(t)\rangle$ is average over $M=10^3$, $10^3$, $10^3$, $10^2$ simulations for $N=12$, 24, 48, 96 respectively.}
\end{figure*}
 
\begin{figure*}
\psfrag{rho}[c]{$\rho(\Delta\eta_i^w)$, $\rho(\Delta\eta_i^l)$}
\psfrag{rhow}[c]{$\rho(\Delta\eta_i^w)$}
\psfrag{rhol}[c]{$\rho(-\Delta\eta_i^l)$}
\psfrag{rhoth}[c]{$\rho(\Delta\eta^{\text{th}})$}
\psfrag{w}{$w$}
\psfrag{l}{$l$}
\psfrag{D eta}[c]{$-\Delta\eta_i^l$, $\Delta\eta_i^w$}
\psfrag{D eta th}[c]{$-\Delta\eta_i^l$, $\Delta\eta_i^w$, $\Delta\eta^{\text{th}}$}
\psfrag{N=4, Nrun=1e5}[c]{(a) $N=4$, $M=10^5$}
\psfrag{N=6, Nrun=1e5}[c]{(b) $N=6$, $M=10^5$, $\langle\Delta\eta_i^w\rangle=0.0727$, $\langle\Delta\eta_i^l\rangle=-0.0660$}
\psfrag{N=12, Nrun=1e5}[c]{(c) $N=12$, $M=10^5$, $\langle\Delta\eta_i^w\rangle=0.0263$, $\langle\Delta\eta_i^l\rangle=-0.0163$}
\psfrag{N=24, Nrun=1e5}[c]{(d) $N=24$, $M=10^5$, $\langle\Delta\eta_i^w\rangle=$, $\langle\Delta\eta_i^l\rangle=...$}
\psfrag{N=48, Nrun=1e3}[c]{(e) $N=48$, $M=10^3$, $\langle\Delta\eta_i^w\rangle=...$, $\langle\Delta\eta_i^l\rangle=...$}
\psfrag{N=96, Nrun=1e3}[c]{(f) $N=96$, $M=10^3$, $\langle\Delta\eta_i^w\rangle=...$, $\langle\Delta\eta_i^l\rangle=...$}
\includegraphics[width=.99\textwidth]{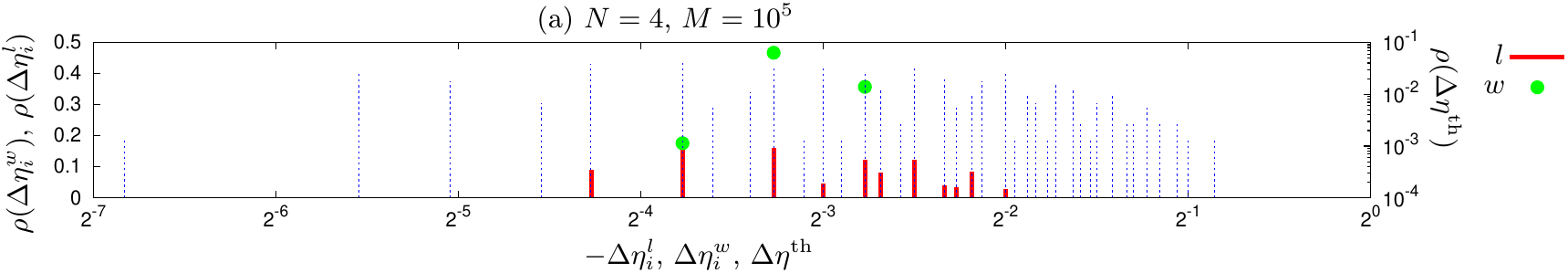}\\[1mm]
\includegraphics[width=.99\textwidth]{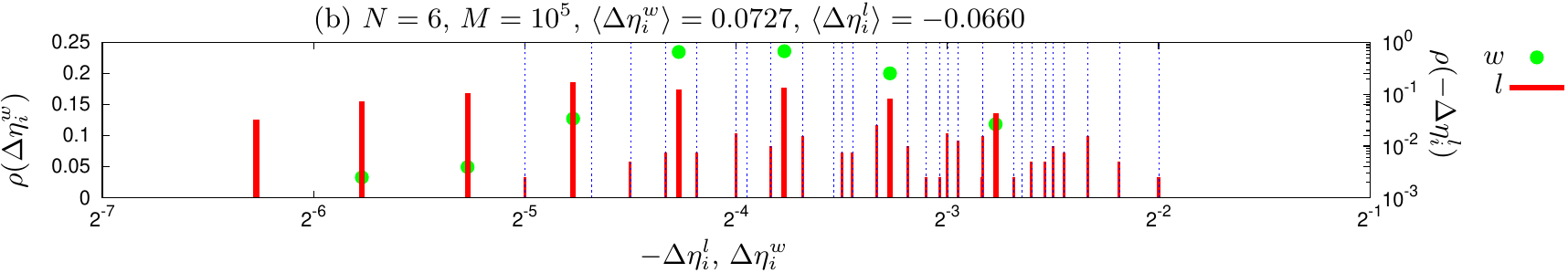}\\[1mm]
\includegraphics[width=.99\textwidth]{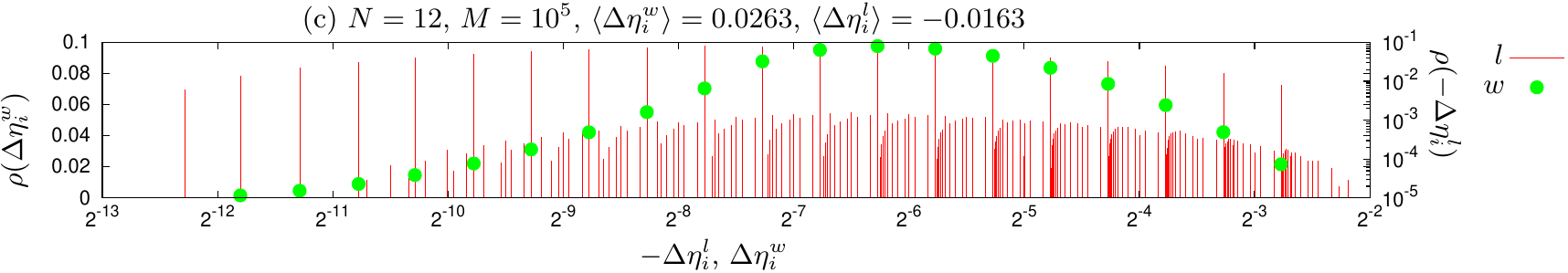}\\[1mm]
\caption{\label{fig:hisdeltaeta}Probability distributions of utility parameter changes during transactions $\Delta\eta_i^{w,l}=\eta_i(t+1)-\eta_i(t)$.
The circles and squares correspond to transactions `winners' ($w$) and `losers' ($l$), respectively.
In sub-figure (a) also available differences between all possible values of $\eta_i$ for $N=4$ agents are marked by blue dashed impulses.
In sub-figure (b) blue dashed impulses indicates positions of $-\Delta\eta_m^l-\Delta\eta_n^l$ ($m\ne n$), where $-\Delta\eta_{m,n}^l$ are values of the first component of $\rho(-\Delta\eta_i^l)$ distribution.
The latter is marked as thick red impulses, while the second component of $\rho(-\Delta\eta_i^l)$ is indicated with thin red lines.
}
\end{figure*}

\begin{figure*}
\psfrag{t}{$t$}
\psfrag{eta_1}{(a) $\eta_1$}
\psfrag{eta_2}{(b) $\eta_2$}
\psfrag{eta_3}{(c) $\eta_3$}
\psfrag{eta_4}{(d) $\eta_4$}
\includegraphics[width=.99\columnwidth]{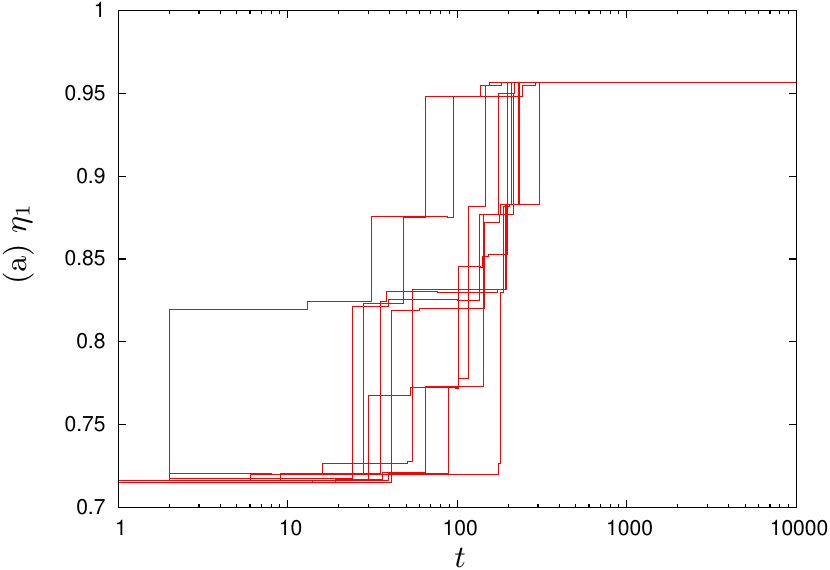}
\includegraphics[width=.99\columnwidth]{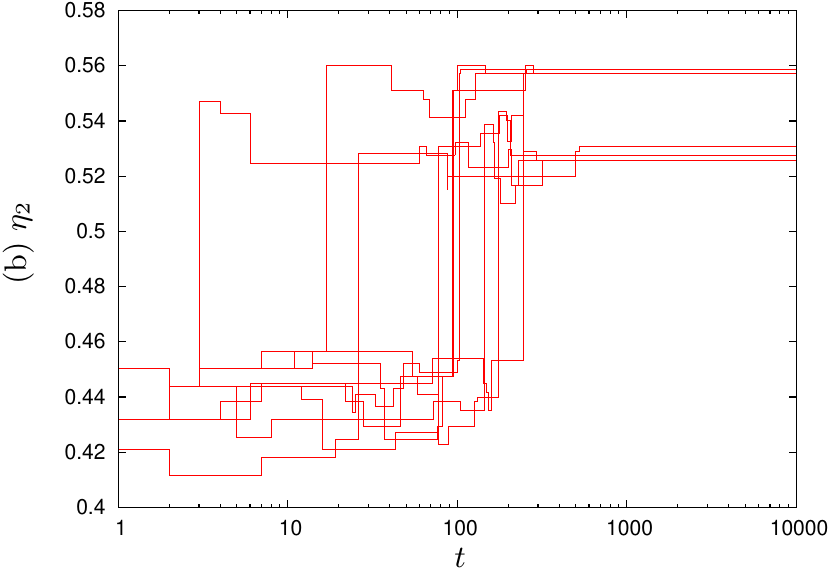}\\
\includegraphics[width=.99\columnwidth]{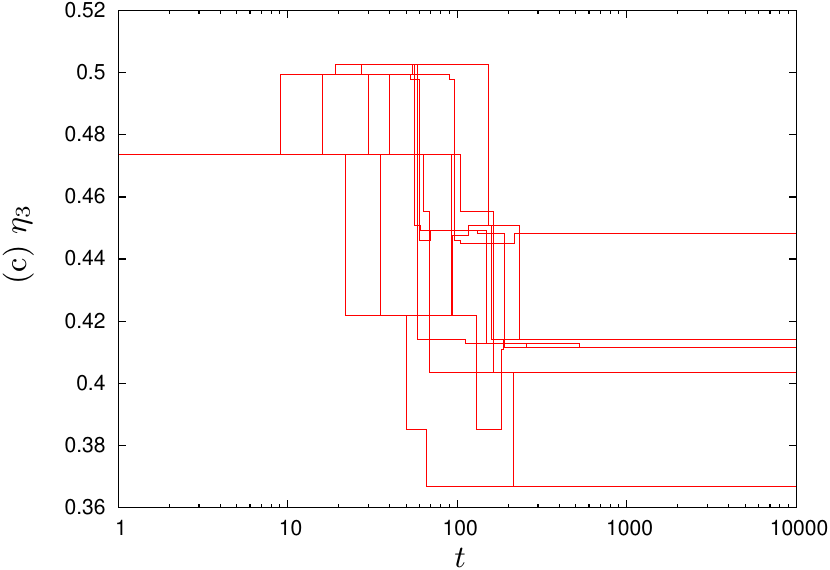}
\includegraphics[width=.99\columnwidth]{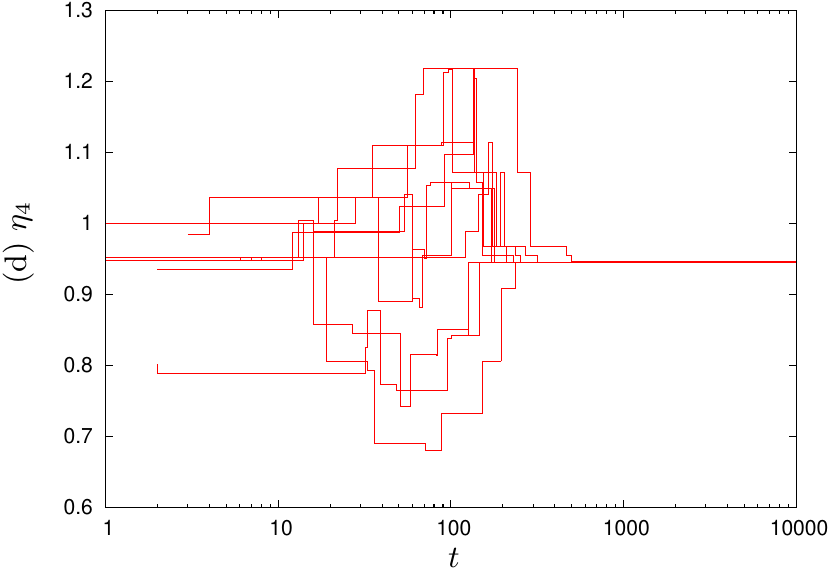}
\caption{\label{fig:initialcond}Starting with exactly the same initial conditions---i.e. identical
{\em i}) sequences of agents goods $(\sigma_1,\sigma_2,\cdots\sigma_N)$,
{\em ii}) agents' priories of goods $\vec\xi_i$ and
{\em iii}) initial order of agents $\vec\zeta_i^{\,0}$---the final hierarchies of agents $\vec\zeta_i^{\,\infty}$ may be different.}
\end{figure*}

\begin{figure*}
\psfrag{rho}{$\rho$}
\psfrag{tau}{$\tau$}
\psfrag{N=12, Nrun=1e3}[c]{(a) $N=12$, $M=10^3$}
\psfrag{N=24, Nrun=1e3}[c]{(b) $N=24$, $M=10^3$}
\includegraphics[width=.45\textwidth]{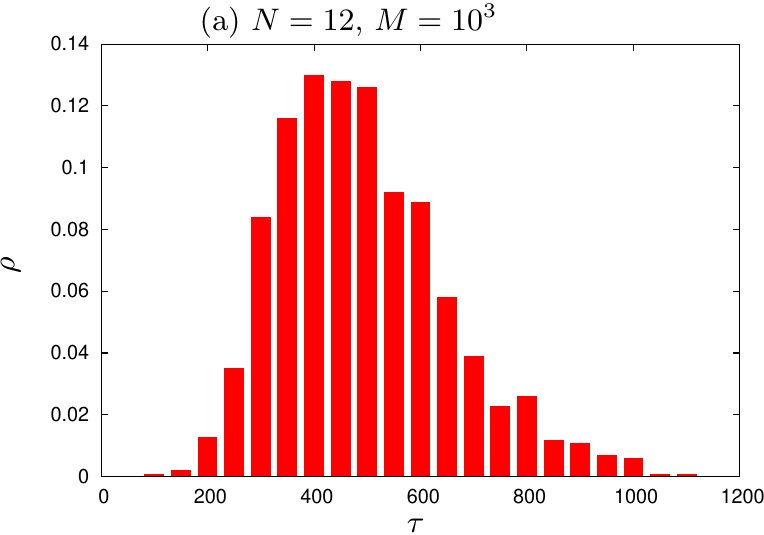}
\includegraphics[width=.45\textwidth]{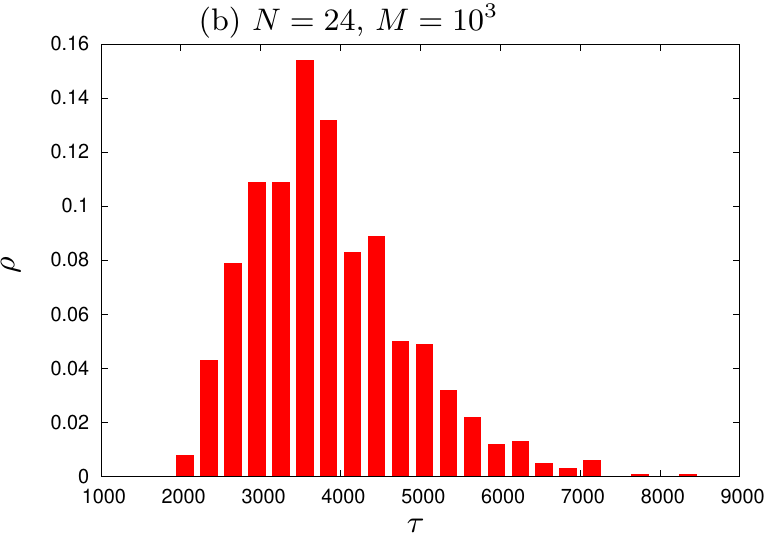}
\caption{\label{fig:histau}The probability distribution of times necessary for reaching a steady state of the system.
The results aver averaged over $M=10^3$ simulations and times $\tau$ data are binned in $\Delta\tau=50$, 300 long classes for $N=12$ and 24, respectively.}
\end{figure*}

\begin{figure}
\psfrag{tau}{$\langle\tau\rangle$}
\psfrag{N}{$N$}
\includegraphics[width=.99\columnwidth]{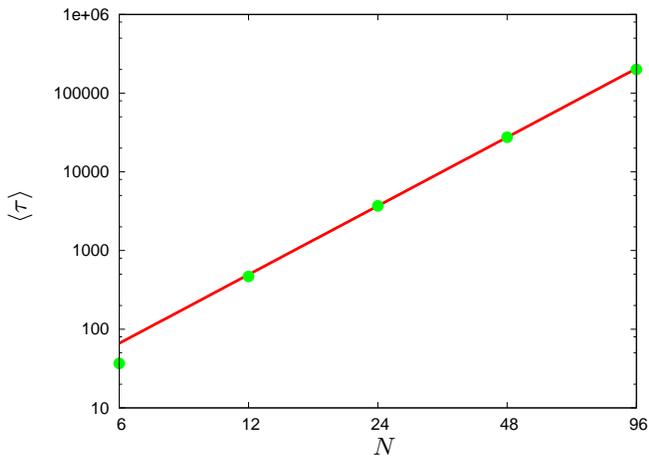}
\caption{\label{fig:tauvsN} An average time $\langle\tau\rangle$ vs. system size $N$.
The data are averaged over $M=10^5$, $10^3$, $10^3$, $10^2$, $10^2$ for $N=6$, 12, 24, 48, 96, respectively.
The linear fit in logarithmic scale indicates dependence $\langle\tau\rangle\propto N^\gamma$, $\gamma=2.901$, $u(\gamma)=0.011$.}
\end{figure}

\begin{figure}
\psfrag{t/Tmax}{$t/\tau_{\text{max}}$}
\psfrag{N=}{$N$}
\psfrag{#links}[c]{$\ell$}
\includegraphics[width=.99\columnwidth]{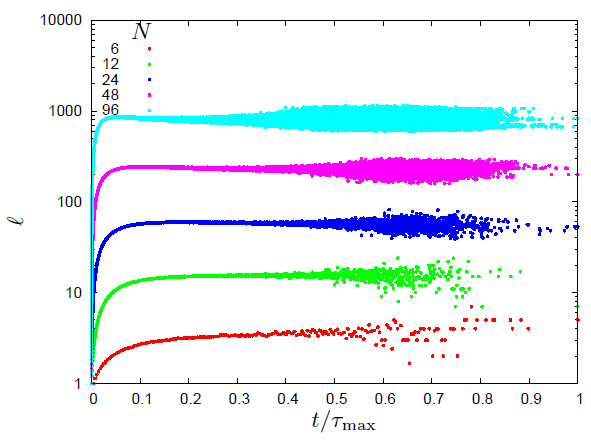}
\caption{\label{fig:numlinks} Number of links $\ell$ among agents for their various number $N$ vs normalised time $t/T_{\text{max}}$.
$M=10^5$, $10^5$, $10^5$, $10^5$, $10^5$ and $\tau_{\text{max}}=275$, $1891$, $11239$, $77486$ and $562925$ for $N=6$, 12, 24, 48, 96, respectively.}
\end{figure}

\begin{figure*}
\psfrag{eta}{$\langle\eta_i\rangle$}
\psfrag{N=6, without}[c]{(a) $N=6$, $\eta_i^{\max}\approx 1.41$}
\psfrag{N=12, without}[c]{(e) $N=12$, $\eta_i^{\max}\approx 1.67$}
\psfrag{N=6, star=1}[c]{(b) $N=6$, $a_*=a_1$}
\psfrag{N=12, star=1}[c]{(f) $N=12$, $a_*=a_1$}
\psfrag{N=6, star=3}[c]{(c) $N=6$, $a_*=a_3$}
\psfrag{N=12, star=4}[c]{(g) $N=12$, $a_*=a_4$}
\psfrag{N=6, star=6}[c]{(d) $N=6$, $a_*=a_6$}
\psfrag{N=12, star=8}[c]{(h) $N=12$, $a_*=a_8$}
\psfrag{t}{$t$}
\psfrag{i}{$i$}
\includegraphics[width=.95\columnwidth]{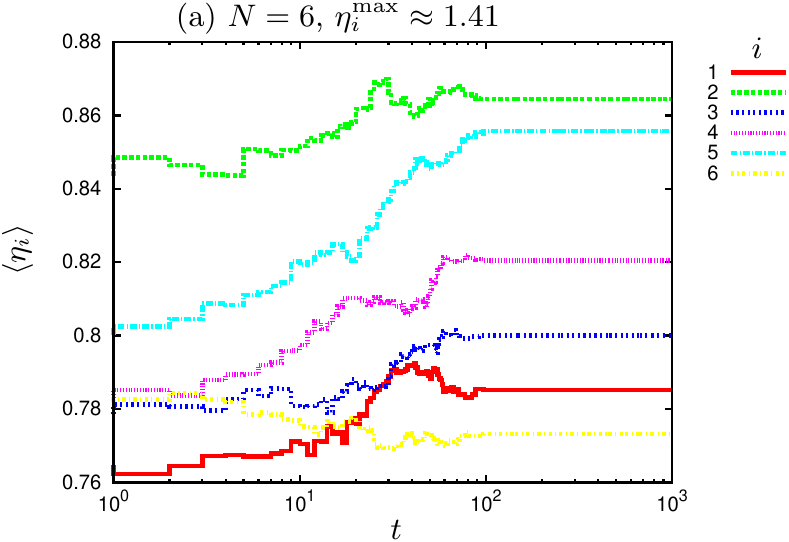}
\includegraphics[width=.95\columnwidth]{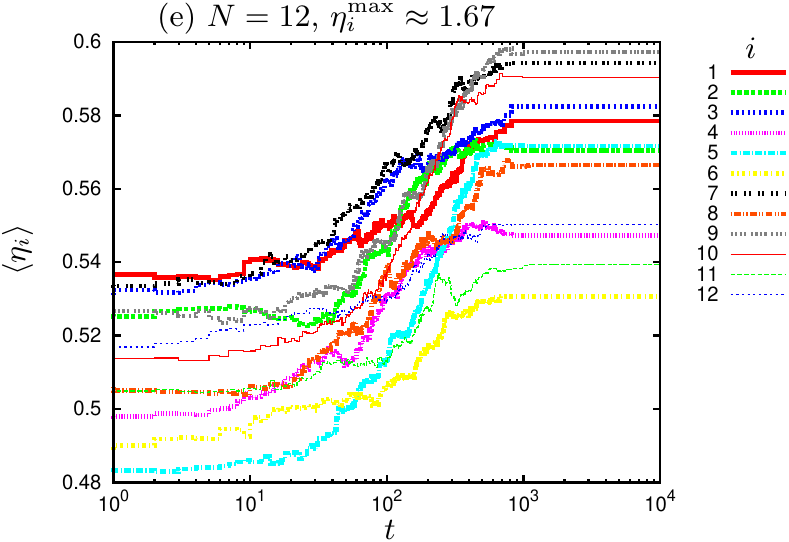}\\
\includegraphics[width=.95\columnwidth]{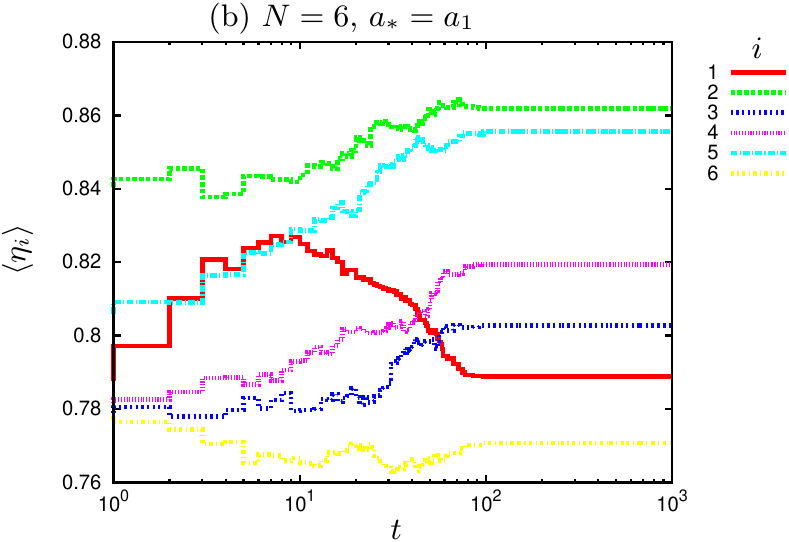}
\includegraphics[width=.95\columnwidth]{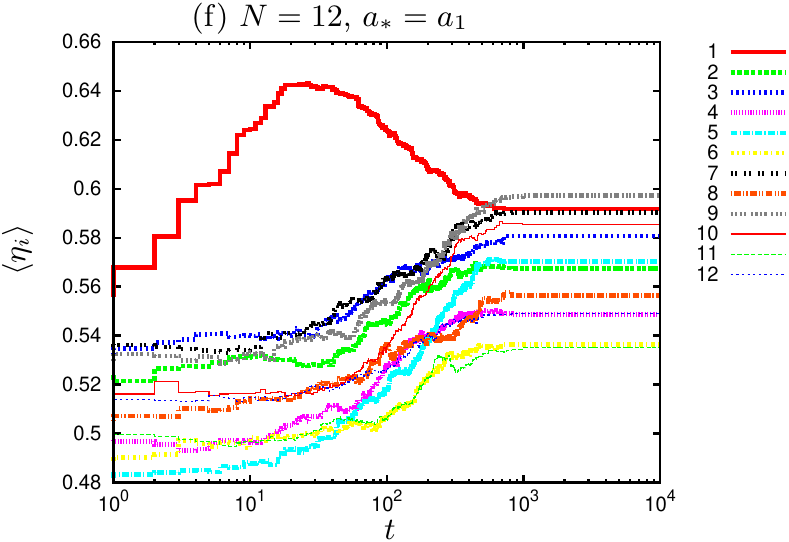}\\
\includegraphics[width=.95\columnwidth]{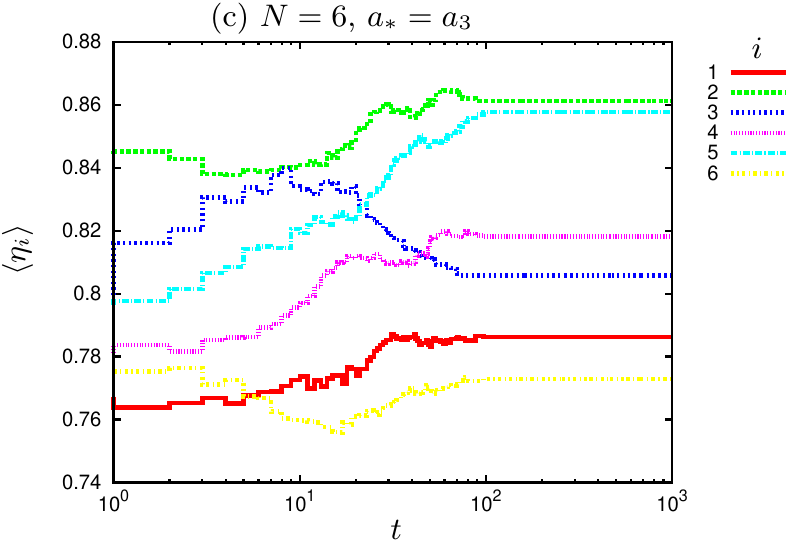}
\includegraphics[width=.95\columnwidth]{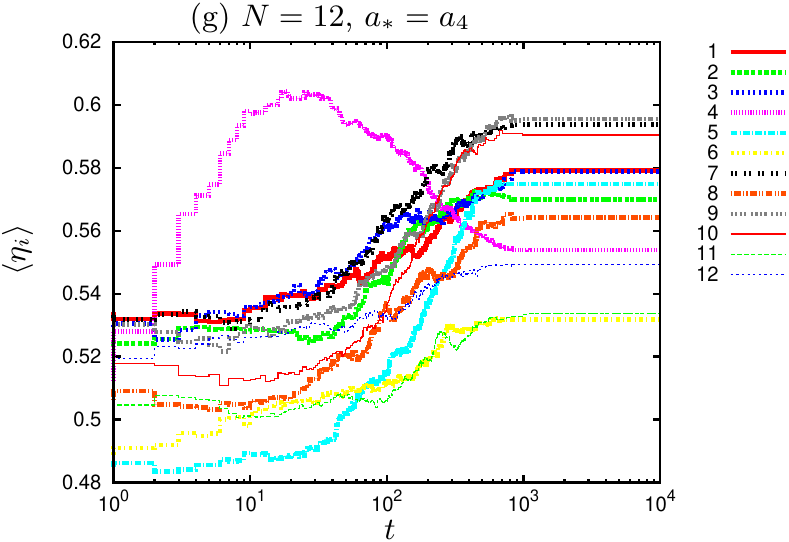}\\
\includegraphics[width=.95\columnwidth]{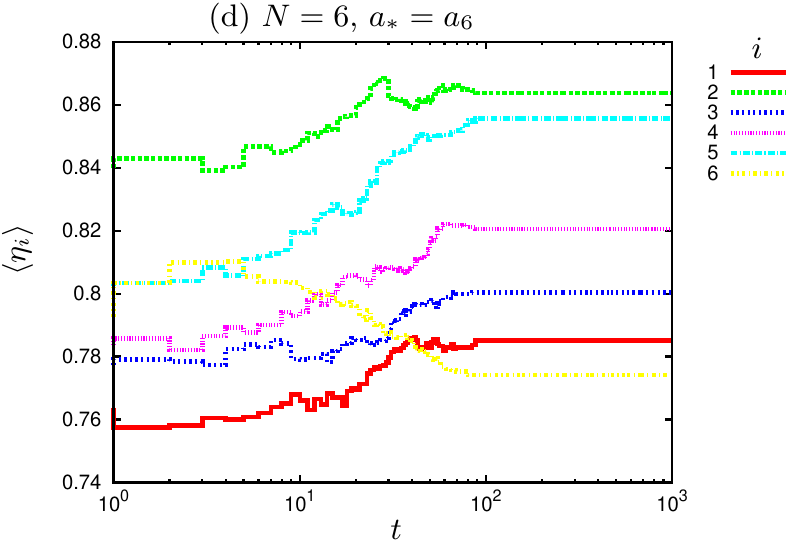}
\includegraphics[width=.95\columnwidth]{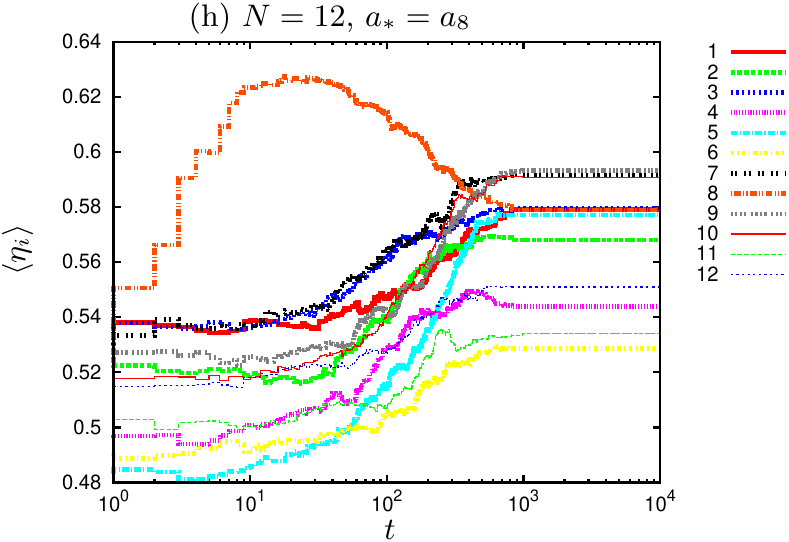}
\caption{\label{fig:privileges}The time evolution of an average parameter $\langle\eta_i(t)\rangle$, where $\langle\cdots\rangle$ is an averaging over $M$ independents simulations.
In sub-figures (a) and (e) agents $a_*$ with extra privileges for making transactions are absent.
}
\end{figure*}

\subsection{\label{sec:kappa}Time evolution of the Kendall's rank correlation coefficient $\kappa$.}

We apply Kendall's rank correlation coefficient to measure the association between {\em an order} $\vec x_i$ of goods in the list $\vec\xi_i$ of $i$-th agent $\vec x_i=(1,2,\cdots,N-1)$ and sequence $\vec y_i=(\nu_{i1},\nu_{i2},\cdots,\nu_{i(N-1)})$ of agent $a_i$'s positions on the lists $\vec\zeta_{k\ne i}^{\,t}$, where $k$ is the label of the agent having good being $j$-th ($j=1,\cdots,N-1$) on $\vec\xi_i$ list.

In Fig.~\ref{fig:kappa} the time evolution of the Kendall's rank correlation coefficients $\kappa_i$ for every agent $i=1,2,\cdots,N$ are presented.
Less than $100$, $10^3$ and $10^4$ time steps are necessary for reaching the steady state for the systems containing $N=6$, 12 and 24 agents, respectively.

In Fig.~\ref{fig:hisdeltakendall} the probability distributions $\rho(\Delta\kappa_i^{w,l})$ of Kendall's statistics changes during transactions $\Delta\kappa_i^{w,l}=\kappa_i(t+1)-\kappa_i(t)$ are presented.
The indexes $w$ and $l$ denote the transactions `winners' and `losers', respectively.
The distribution of $\Delta\kappa_i$ for transactions `winners' $\rho(\Delta\kappa_i^w)$ are slightly shifted towards positive values of $\Delta\kappa_i$ in respect to these distributions for transactions `losers' $\rho(\Delta\kappa_i^l)$.
It is worth to mention that changes in Kendall's statistics $\kappa_i$ do not distinguish between transaction results, i.e. these differences may be positive for the `losers' of transactions and negative for the `winners' of transactions.

\subsection{\label{sec:eta}Properties of $\eta$ utility coefficient.}

In Fig.~\ref{fig:etavst} a time evolution of the utility correlation parameters $\eta_i(t)$ for $N=6$, 12 and 24 agents is presented.
The changes in $\eta_i(t)$ mark times $t$ where transactions take place.
The increase (decrease) of $\eta_i$ parameter is a signature of winning (loosing) of the transaction by the agent $a_i$.

In Fig.~\ref{fig:kappavseta} the comparison of time evolution of parameters $\kappa_i(t)$ and $\eta_i(t)$ for two agents $a_1$ and $a_2$ among group of a dozen of agents are presented.
As one can see, the increase of $\kappa_i$ in time $t$ may by accompanied by decrease of $\eta_i$ and {\em vice versa}.
Moreover, sometimes the changes of $\eta_i$ are not associated with corresponding changes of $\kappa_i$ [see for instance jumps of $\eta_1(t)$ for $t\approx 250$ and $t\approx 400$ when $\kappa_1(t)$ is constant, as shown in Fig.~\ref{fig:kappavseta}(a)].

In Fig.~\ref{fig:aveta} the time evolution of an average parameter $\eta\equiv\bar\eta_i$ for various system sizes $N$ are presented.
A bar over symbol $\eta_i$ stands for an average over all $N$ agents.
Different green dotted curves represent various trajectories of $\eta(t)$ for five among $M$ various simulations.
The thick red solid line represents additional average $\langle\cdots\rangle$ over $M$ simulations with $M=10^3$, $10^3$, $10^3$ and $10^2$ for $N=12$, 24, 48 and 96, respectively.
Although a~single agent may loose his/her `prestige' during system evolution---e.g. $[\eta_i(t\to\infty)-\eta_i(t=0)]<0$, as for agents $a_2$ and $a_4$ in Fig.~\ref{fig:etavst}(a)---and even temporally the group of agents may loose their average `prestige' $\eta(t)$ the difference $[\eta(t\to\infty)-\eta(t=0)]$ is positive.
Moreover, various simulation average out these local loses of the group prestige $\eta(t)$ and $\langle\eta(t)\rangle$ grows monotonically with $t$.
The latter means that, in average, the individual loses of the `prestige' $[\eta_i(t+1)-\eta_i(t)<0]$ by the transaction `losers' $a_i$ are globally compensated by an increase of `prestige' of the transaction `winners'.

In order to quantitatively measure the average advantage in transactions results between transaction `winners' ($w$) and `losers' ($l$) we check the probability distributions of utility parameter changes during transactions $\Delta\eta_i^{w,l}=\eta_i(t+1)-\eta_i(t)$.
In Fig.~\ref{fig:hisdeltaeta} these distributions are presented for various system sizes $N=4$, 6 and 12.
The results are averaged over $M=10^5$ independent simulations.
As the results are presented in semi-logarithmic scale the values $\rho(\Delta\eta_i=0)$ are omitted, but $\rho(\Delta\eta_i=0)>0$ occurs only for a small system sizes and vanishes for $N=12$ and larger.

As we mentioned earlier in Sec.~\ref{sec:kappa} and presented in Fig.~\ref{fig:hisdeltakendall}, the sign of changes $\Delta\kappa_i$ is not able to distinguish among the transactions `winners' and `losers'.
In contrast, $\Delta\eta_i$ are positive for transaction `winners' and they are negative for transaction `losers'.
The absolute values of averages for these distributions are larger for transaction `winners' than for transaction `losers': $\langle-\Delta\bar\eta_i^l\rangle < \langle\Delta\bar\eta_i^w\rangle$, where a bar over symbols stands for an average over all $N$ agents while $\langle\cdots\rangle$ indicates an average over $M$ independent simulations.
These averages $\langle\Delta\bar\eta_i^{w,l}\rangle$ are indicated in Figs.~\ref{fig:hisdeltaeta}(a)-(c) headlines.

The probability distribution $\rho(\Delta\eta_i^w)$ for transaction `winners' are smooth and equidistant (in logarithmic-scale).
The same distribution for the transaction `losers' $\rho(-\Delta\eta_i^l)$ may be split into two parts.
The first component of $\rho(-\Delta\eta_i^l)$ is quite similar to $\rho(\Delta\eta_i^w)$ except of mentioned earlier small shift of its average towards lower values.
This component is assiociated with transactions similar to those presented in Eq.~\eqref{eq:t=2}, i.e. when two differnt agents $(a_{l_1},a_{l_2\ne l_1})$ stay a head of transaction `winners' ($a_{w_1},a_{w_2}$) on $\zeta^t_{w_1}$ and $\zeta^t_{w_2}$ lists. 
The second part of $\rho(-\Delta\eta_i^l)$ is composed with plenty but seldom observed changes $\Delta\eta_i^l$ occurring for $N=12$ three orders of magnitude less often than a mode of this distributions.
This part of $\rho(-\Delta\eta_i^l)$ is associated with transactions with `double-losers' [see Eq.~\eqref{eq:t=4}], when agent $a_{l}$ is directly before `winners' ($a_{w_1},a_{w_2}$) on $\zeta^t_{w_1}$ and $\zeta^t_{w_2}$ lists.
In such case $\Delta\eta_i^l$ is a result of two demoting operations.
In Fig.~\ref{fig:hisdeltaeta}(b) blue dashed impulses indicates positions of $\Delta\eta_m^l+\Delta\eta_n^l$ ($m\ne n$), where $\Delta\eta_{m,n}^l$ are values of the first component of $\rho(\Delta\eta_i^l)$ distribution.
And indeed, these positions perfectly match observed values of $\Delta\eta_i^l$ in the second component of $\rho(\Delta\eta_i^l)$ distribution.

To check how possible differences $\Delta\eta_i^l$ are realized during transaction we checked the distribution of differences of two sums 
\begin{equation}
\sum_{k=1}^3 2^{-(k+\nu_k)/2}-\sum_{m=1}^3 2^{-(m+\mu_m)/2}\equiv\Delta\eta^{\text{th}},
\label{eq:sum-sum}
\end{equation}
where $\mu_m,\nu_k=1$, 2, 3, what corresponds to possible differences in $\eta_i$ in the group of $N=4$ agents.
The sums in Eq.~\eqref{eq:sum-sum} may be composed in $3^3$ ways\footnote{Please note, that for a group of $N=4$ agents a number $3^3=27$ is also a number of possible sequences $\nu_{ij}$, which appears in $\eta_i$ parameter definition [Eq.~\eqref{eq:etai}].} what yields $27^2$ possible differences $\Delta\eta^{\text{th}}$.
Of course, values of $\Delta\eta^{\text{th}}$ may be identical for various sequences $\mu_{m=1,2,3}$ and $\nu_{k=1,2,3}$.
The values of $\rho(\Delta\eta_i^{\text{th}})$ are presented on the right axis of Fig.~\ref{fig:hisdeltaeta}(a).
The half of probability distribution $\rho(\Delta\eta_i^{\text{th}})$ is marked in Fig.~\ref{fig:hisdeltaeta} as a dashed blue impulses, except of value $\rho(\Delta\eta_i^{\text{th}}=0)=\frac{45}{729}$.
The observed changes $\Delta\eta_i^{w,l}$ agree with their predicted positions $\Delta\eta_i^{\text{th}}$, however, only small fraction of available differences $\Delta\eta_i^{\text{th}}$ is realised during transactions.

In Fig.~\ref{fig:initialcond} again the time evolution of $\eta_i(t)$ parameter is presented.
The $M=10$ evolution of $\eta_i(t)$ on every sub-figure \ref{fig:initialcond}(a)-(d) correspond to the same agent $a_i$ ($i=1,\dots,4$).
In all cases the initial conditions---i.e.
{\em i}) sequences of agents goods $(\sigma_1,\sigma_2,\cdots\sigma_N)$, 
{\em ii}) agents' priorities of goods $\vec\xi_i$ and 
{\em iii}) initial order of agents $\vec\zeta_i^{\,0}$---are identical.
The only difference among these simulations is a sequence of agents pairs $(a_m,a_n)$ selected in subsequent time steps $t$.
With the same initial conditions the evolution of $\eta_i(t)$ may lead to identical final sate [see Figs.~\ref{fig:initialcond}(a) and (d)] or not [see Figs.~\ref{fig:initialcond}(b) and (c)].
But, in principle, the final state of the whole system will be different for different simulations...

\subsection{\label{sec:tau}Distribution of times necessary for reaching a steady state of the system.} 

In Fig.~\ref{fig:histau} the probability distribution of times necessary for reaching a steady state of the system for various system sizes $N$ are presented.
The results aver averaged over $M=10^5$, $10^3$, $10^3$ simulations and times $\tau$ data are binned in $\Delta\tau=1$, 50, 300, long classes for $N=6$, 12 and 24, respectively.
For the smallest simulated system ($N=6$) plenty simulations do not contain any transaction and thus $\rho(0)>0$. 

In Fig.~\ref{fig:tauvsN} an average time $\langle\tau\rangle$ of reaching the steady state of the system for various system sizes $N$ is presented.
The $\langle\cdots\rangle$ symbol stands for an averaged over $M=10^5$, $10^3$, $10^3$, $10^2$, $10^2$ for $N=6$, 12, 24, 48, 96, respectively.
The least squares fit in logarithmic scale indicates dependence $\langle\tau\rangle\propto N^\gamma$ with $\gamma=2.901$ and its uncertainty $u(\gamma)=0.011$.

\subsection{\label{sec:net}Network of promoted agents.}

In Fig.~\ref{fig:numlinks} the time evolution of a number $\ell$ of links among $N$ agents---i.e. the half of the number of unities in an adjacency matrix which define the promoted agents network topology---is presented.
The adjacency matrix is a symmetric binary matrix, which elements are zero or one ($a_{mn}\in\{0,1\}$, $a_{mn}=a_{nm}$)~\cite{Wilson}.
The nonzero element of the adjacency matrix $a_{mn}$ indicates that a link among agent $a_m$ and $a_n$ exists.
Initially, the adjacency matrix does not contain any unities: $a_{mn}=0$ for $m,n\in\{1,2,\cdots, N\}$.
Every time, when transaction among agents $a_m$ and $a_n$ takes place, we set $a_{mn}=a_{nm}=1$ and $a_{l_1l_2}=a_{l_2l_1}=0$, where $l_{1,2}$ are the labels of transaction's `losers' ($a_{l_1}$ and $a_{l_2}$).
The results are averaged over $M=10^5$ simulations.
As we mentioned in Sec.~\ref{sec:tau} the average times $\langle\tau\rangle$ of reaching a steady state of the systems scales roughly as $N^{\gamma}$ and thus we present time evolution of $\ell$ vs. normalised time $t/\tau_{\text{max}}$, where 
\[
\tau_{\text{max}}=\max_{1\le k\le M}(\tau_k(N))
\]
is the longest time $\tau_k(N)$ of reaching a steady state in $M$ simulations for fixed value of $N$.
These times are $\tau_{\text{max}}=275$, $1891$, $11239$, $77486$ and $562925$ for $N=6$, 12, 24, 48, 96, respectively.

As shown in Fig. ~\ref{fig:numlinks}, at the initial stage the number of links increases. Once more and more transactions are done, some links marking the previous transactions are erased. The mean number of links in the stationary state is roughly proportional to the system size. However, the plot $\ell(t)$ shows a flat maximum. This is a remarkable result, because if the links are selected and erased randomly, the time dependence of their number is known to increase monotonously till a saturation.
Apparently, some kind of short-range order of selected links is active here: as the number of links increases, the links between the losers and the winners are preferably created between actors who did not cooperate yet. After a time, however, the short-range order is destroyed, and the stationary density of links is slightly lower than its maximal value.

\subsection{\label{sec:privileges}The role of agents privileges.}

According to Ref.~\cite{Crozier}, information plays a key role in the distribution of power. In particular, the actual state of transactions performed by an actor may remain hidden for other actors, and therefore may be an important part of the related sphere of uncertainty. To check how is works in our simulation, we allow one actor to be completely informed on the state of the whole system. This actor does not need to wait for the moment when by chance he meets another actor with whom he may decide to make a transaction. Instead, he can act at every moment and therefore he has an effective priority to perform all transactions which are productive for both contarctors. Namely, every time $t$ when transaction among randomly selected agents $(a_m,a_n)$ takes place, this special agent $a_*$ is invited for taking part in some extra transactions. These additional transactions are performed until further agent's $a_*$ promotions on any of $\vec\zeta^t_i$ lists ($i=1,\cdots,N$ and $i\ne *$) are not possible.

In Fig.~\ref{fig:privileges} the time evolution of utility parameters $\langle\eta_i(t)\rangle$ for several agents in $N=6$ and $N=12$ large groups are presented.
The symbol $\langle\cdots\rangle$ marks an averaging procedure over $M$ independent simulations.
In Figs.~\ref{fig:privileges}(a) and~\ref{fig:privileges}(e) the situations corresponding to the absence of $a_*$ are presented.
Also maximal available values of $\eta_i^{\max}$ for $N=6$ and 12 are printed there.
The agents $a_*$ priviledged to take part in these extra transactions are indicated in the sub-figues headlines.

The privilege of additional transaction seems to have only temporary effect on value of $\eta_*$, which grows in time much faster than $\eta_{i\ne *}$.
In group of dozen agents $N=12$ the increase of $\eta_*$ is observed until $t\approx 20$ time steps.
After this time the additional transactions are not observed and agent $a_*$ takes part in transaction only as a loser.
The ultimate values $\eta_*(t\to\infty)$ are not much higer than values observed without these additional transactions [cf. Figs.~\ref{fig:privileges}(a) and~\ref{fig:privileges}(e)].
Please note, that also largest value of the utility parameter
\[
\max_t(\eta_*(t))
\]
reached during simulation is definitely lower than its maximal available value $\eta_i^{\max}$.
The efect of additional transactions manifest itself much weaker in the smaller groups, and is almost invisible for $N=6$.

It seems natural to state that the final state when all profitable transactions are done can be interpreted as the state of full information of all actors. Within these terms, a nice conclusion of this part of calculations is that the advantage of being informed works only when other actors remain uninformed. 

\section{\label{sec:disc}Discussion and conclusions.}

Summarizing our main results, we can state at first that the time of simulation is finite. The state at the end of calculations is well defined: there is not a single pair of agents which can carry out a mutually successful transaction.
When the time of evolution is averaged over a number of simulations, its mean value varies as a power law with the number $N$ of agents. Further, the final state depends not only on the initial order of all lists, but also on the order of pairs which carry out their transactions. Once this order is prescribed, the system evolution is fully deterministic. Within our interpretation, this order is determined by autonomous decisions of the actors. If this is accepted, our numerical results indicate that the final state of the system does depend on these decisions. The latter conclusion is consistent with the postulate of spheres of uncertainty, controlled by the actors \cite{Crozier}.

With the results listed above, we gain an analogy between the characteristics of the final state and the relations of power; recall the key role of the latter in the writings of Crozier and Friedberg \cite{Crozier}. There, we find statements that the power is a feature of a relation, and not of the actors. Further, according to Ref.~\cite{Crozier}, the power is intransitive, the power relations are not equivalent, and the power can be identified as a difference in profits of a cooperation. All these statements find their counterparts in our formulation and our numerical results. In particular, if an actor A is at the first place in the list of an actor B, and the actor B is not at the first place in the list of an actor A, this means that the actor B would like to be advanced in the list of A, but he has nothing to offer. Accordingly, as we read in Ref.~\cite{Crozier} the power is an opportunity {\em not} to fulfill the partner's wish.

Another result is that the time dependence of the Kendall coefficients for individual actors is not monotonous, and the same can be stated about the coefficient $\eta_i$. Recall that the Kendall coefficient is a measure of the accordance of the individual rank of goods with what can be interpreted as an access to those goods. For an extremely successfull actor, who has number one in the lists of all agents, the Kendall coefficient is expected to be close to zero. Then, the role of a measure of success is played by the utility parameter $\eta_i$, introduced as a weighted sum of accesses to the goods; the weight is determined by the number of the good at the actor's list. It is interesting that the value of the utility parameter, averaged over all actors, does increase; the numerical evaluation gives roughly the same increase $\Delta \eta$ close to 0,04 for any system size. This can be interpreted as an `optimistic' statement that the collusions are profitable for the whole system. However, we stress that the 
profits are directed only to the members of the organization, what is not necessarily desirable for, say, the national health service or a state university.

\begin{acknowledgments}
This research was supported in part by PL-Grid Infrastructure and Polish Ministry of Science and Higher Education.
\end{acknowledgments}


\begin{thebibliography}{9}
\bibitem{Crozier} M. Crozier and E. Friedberg, {\it L’acteur et le systeme. Les contraints de l'action collective}, Editions du Seuil, Paris 1977 ({\it Actors and systems. The politics of collective action}, Chicago UP, Chicago 1980). 
\bibitem{ap1} C. Barmeyer and U. Mayrhofer, \href{http://dx.doi.org/10.1108/IJOA-06-2013-0676}{\it How has the French context shaped the organization of the Airbus group?}, International Journal of Organizational Analysis {\bf 22}(4) 426--448 (2014).
\bibitem{ap2} M. Geppert, K. Williams and M. Wortmann, \href{http://dx.doi.org/10.1177/0959680114544015}{\it Micro-political game playing in Lidl: A comparison of store-level employment relations}, European Journal of Industrial Relations {\bf 21}(3) 241--257 (2015).
\bibitem{3u} I. Bleiklie, J. Enders and B. Lepori, \href{http://dx.doi.org/10.1177/0170840615571960}{\it Organizations as penetrated hierarchies: Environmental pressures and control in professional organizations}, Organization Studies {\bf 36}(7) 873--896 (2015).
\bibitem{va1} K. Vafa\"i, \href{http://dx.doi.org/10.1016/j.ejpoleco.2004.07.006}{\it Abuse of authority and collusions in organizations}, European Journal of Political Economy {\bf 21}(2) 385--405 (2005).
\bibitem{va2} K. Vafa\"i, {\it Information in Hierarchies}, Documents de travail du Centre d'Economie de la Sorbonne 2012.86, ISSN: 1955-611X (2012).
\bibitem{va3} K. Vafa\"i, {\it Supervision in Firms}, Documents de travail du Centre d'Economie de la Sorbonne 2012.84, ISSN: 1955-611X (2012).
\bibitem{s1} C. Sibertin-Blanc, F. Amblard and M. Mailliard, \href{http://dx.doi.org/10.1007/11775331_1}{\it A coordination framework based on the Sociology of Organized Action}, Lecture Notes in Computer Science {\bf 3913} 3--17 (2006).
\bibitem{s2} J. El-Gemayel, C. Sibertin-Blanc and P. Chapron, \href{http://dx.doi.org/10.1007/978-3-642-16098-1_18}{\it Impact of tenacity upon the behaviors of social actors},
Studies in Computational Intelligence {\bf 325} 287--306 (2011).
\bibitem{s3} C. Sibertin-Blanc, P. Roggero, F. Adreit, B. Baldet, P. Chapron, J. El-Gemayel, M. Mailliard and S. Sandri, \href{http://dx.doi.org/10.18564/jasss.2278}{\it SocLab: A framework for the modeling, simulation and analysis of power in social organizations}, JASSS---The Journal of Artificial Societies and Social Simulation {\bf 16}(4) 8 (2013).
\bibitem{mw} \url{http://www.merriam-webster.com/}
\bibitem{wolf} S. Wolfram, \href{https://www.wolframscience.com/nksonline/toc.html}{\em A new kind of science}, Wolfram Media, Inc., Champaign (IL), 2002.
\bibitem{Kendall} M.~Kendall, \href{http://dx.doi.org/10.1093/biomet/30.1-2.81}{\it A new measure of rank correlation}, Biometrika {\bf 30}(1--2) 81--89 (1938).
\bibitem{Wilson} R.~J.~Wilson, {\em Introduction to graph theory} (5th ed.), Pearson, 2012.
\end{thebibliography}
\end{document}